\documentclass{article}
\usepackage[T1]{fontenc} 
\usepackage[utf8]{inputenc} 
\usepackage{ismir,amsmath,cite,url}
\usepackage{graphicx}
\usepackage{color}

\usepackage{booktabs}

\usepackage{lineno}

\usepackage{tikz}
\usepackage{graphicx}
\usetikzlibrary{positioning}

\usepackage[symbol]{footmisc}
\renewcommand{\thefootnote}{\fnsymbol{footnote}}
\title{DarkGAN: Exploiting Knowledge Distillation for Comprehensible Audio Synthesis with GANs}

\oneauthor
  {Javier Nistal,$^{1, 2}$ Stefan Lattner,$^2$ Ga\"el Richard$^1$} {$^1$LTCI, Telecom Paris, IP Paris, France\\ $^2$Sony Computer Science Laboratories (CSL), Paris, France\\ 
  }
  
\def\authorname{J. Nistal, S. Lattner, and G. Richard}

\begin{document}

\maketitle
\begin{abstract}
Generative Adversarial Networks (GANs) have achieved excellent audio synthesis quality in the last years. However, making them operable with semantically meaningful controls remains an open challenge. An obvious approach is to control the GAN by conditioning it on metadata contained in audio datasets. Unfortunately, audio datasets often lack the desired annotations, especially in the musical domain.
A way to circumvent this lack of annotations is to generate them, for example, with an automatic audio-tagging system.
The output probabilities of such systems (so-called "soft labels") carry rich information about the characteristics of the respective audios and can be used to distill the knowledge from a teacher model into a student model. In this work, we perform knowledge distillation from a large audio tagging system into an adversarial audio synthesizer that we call DarkGAN.
Results show that DarkGAN can synthesize musical audio with acceptable quality and exhibits moderate attribute control even with out-of-distribution input conditioning. We release the code and provide audio examples on the accompanying website.

\end{abstract}
\section{Introduction}\label{sec:introduction}

Generative Adversarial Networks (GANs) \cite{Goodfellow2013} have achieved impressive results in image and audio synthesis \cite{DBLP:conf/cvpr/KarrasLAHLA20, DBLP:conf/iclr/BrockDS19, DBLP:conf/cvpr/Park0WZ19, gansynth, nistal2}. However, it is still an open challenge to learn comprehensible features that capture semantically meaningful properties of the data.
In the graphical domain, semantic control is achieved with GANs using semantic layouts \cite{DBLP:conf/cvpr/Park0WZ19} or high-level attributes learned through unsupervised methods \cite{DBLP:conf/cvpr/KarrasLAHLA20}. Other works achieve disentanglement through regularization terms \cite{DBLP:conf/eccv/PeeblesPZE020} or explore the latent space for human-interpretable factors of variation \cite{DBLP:conf/icml/VoynovB20, DBLP:conf/cvpr/ShenGTZ20}. The great success of these approaches is partly enabled by the availability of large-scale image datasets containing rich semantic annotations \cite{imagenet, DBLP:conf/cvpr/CaesarUF18, DBLP:journals/corr/abs-1708-07747}. However, the context is different in the audio domain, where datasets are scarce and often limited in size and availability of annotations.

Therefore, in this work, we test if limited annotations in audio datasets can be circumvented by taking a Knowledge Distillation (KD) approach. To that end, we utilize the soft labels generated by a pre-trained audio-tagging system for conditioning a GAN in an audio generation task. More precisely, we train the GAN on a subset of the NSynth dataset \cite{wavenetae}, which contains a wide range of instruments from acoustic, electronic, and synthetic sources. For that dataset we generate soft labels with a publicly available audio-tagging model \cite{DBLP:journals/taslp/KongCIWWP20}, pre-trained with attributes of the AudioSet ontology \cite{ DBLP:conf/icassp/GemmekeEFJLMPR17}. This ontology contains a structured collection of sound events from many different sources and descriptions of around 600 attributes obtained from YouTube videos (e.g., "singing bowl", "sonar", "car", "siren", or "bird").

The soft labels indicate how much of the different characteristics are contained in a specific sound (e.g., a synthesizer sound may have some similarity with a singing bowl or a sonar pulse). We hope that the generative model can distill such characteristics (e.g., the "essence" of a singing bowl sound) and is then able to emphasize them in the generation. The slight similarities to specific categories in data that can be distilled using soft labels were coined "Dark Knowledge" in \cite{DBLP:journals/corr/HintonVD15}. Therefore, we call the proposed model DarkGAN.


This paper introduces a generic audio cross-task KD framework for transferring semantically meaningful features into a neural audio synthesizer. We implement this framework in DarkGAN, an adversarial audio synthesizer for comprehensible and controllable audio synthesis. We perform an experimental evaluation on the quality of the generated material and the semantic consistency of the learned attribute controls. Numerous audio examples are provided in the accompanying web page,\footnote{https://an-1673.github.io/DarkGAN.io/} and the code is released for reproducibility.\footnote{https://github.com/SonyCSLParis/DarkGAN}


In what follows, we first mention relevant state-of-the-art works in neural audio synthesis and KD (see Sec.~\ref{sec:previous}). In Sec.~\ref{sec:background}, some background on dark knowledge and KD is given, and its application to controllable neural audio synthesis is motivated. Next, we describe the experimental framework of DarkGAN (see Sec.~\ref{sec:experiments}). In Sec.~\ref{sec:discussion} we provide a discussion of the results, and conclude in Sec.~\ref{sec:conclusion}.

\section{Previous work}\label{sec:previous}
In this section we review some of the most important works on neural audio synthesis and knowledge distillation, paying particular attention to those works tackling tasks similar to ours.

\vspace{-0.2cm}
\subsection{Neural Audio Synthesis}
\vspace{-0.2cm}

Many works have applied deep generative methods to address general audio synthesis. These can be categorised into \emph{exact}, \emph{approximate}, and \emph{implicit} density estimation methods. In the first category, autoregressive models of raw audio are state-of-the-art in different audio synthesis tasks \cite{wavenet, wavenetae, melnet}. Popular \emph{approximate} density estimation methods are based on Variational Auto-Encoders (VAE) \cite{vae}. One of the main advantages of VAEs compared to other approaches is the control they offer over the generative process by manipulating a latent space learned directly from the audio data \cite{planetdrums}. Even though latent spaces tend to self-organize according to high-level dependencies in the data, these are still difficult to interpret. Therefore, some works try to impose musically meaningful priors over the structure of these spaces \cite{flowsynth, EslingCB18, Philippe}, or enforce an information bottle-neck by restricting such latent codes to discrete representations to capture fundamental and meaningful features \cite{jukebox, DBLP:journals/corr/abs-2102-05749}. 

Generative Adversarial Networks (GANs) \cite{Goodfellow2013} belong to the \emph{implicit} density estimation methods. Applications of GANs to audio synthesis have mainly focused on speech tasks \cite{SaitoTS18, Kaneko, timbertron, DBLP:conf/iclr/BinkowskiDDCECC20, DBLP:conf/nips/KongKB20, melgan, DBLP:conf/icassp/YamamotoSK20}. The first application to synthesis of musical audio was WaveGAN \cite{wavegan}. Although it did not match autoregressive baselines such as WaveNet \cite{wavenet} in terms of audio quality, it could generate piano and drum sounds quickly and in an entirely unconditional way. Recent improvements in the stabilization and training of GANs \cite{Karras2017, Gulrajani2017, SalimansGZCRCC16} enabled  GANSynth \cite{gansynth} to outperform WaveNet baselines on the task of audio synthesis of musical notes using sparse, pitch conditioning labels. Follow-up works building on GANSynth applied similar architectures to conditional drum sound synthesis using different metadata \cite{nistal2, Drysdale2020ADVERSARIALSO}. DrumGAN \cite{nistal2} synthesizes a variety of drum sounds based on high-level input features describing timbre (e.g., boominess, roughness, sharpness). A few other works have used GANs in a variety of audio tasks like Mel-spectrogram inversion \cite{melgan}, audio domain adaptation \cite{Hosseini, MichelsantiT17} or audio enhancement \cite{DBLP:conf/icassp/BiswasJ20}.


\subsection{Knowledge Distillation}\label{sec:kd}

High-performing models are often built upon classifier ensembles that aggregate their predictions to improve the overall accuracy. Despite having excellent performance, these models tend to be large and slow, impeding their use in memory-limited and real-time environments. Different methods exist for optimizing memory consumption and reducing the size of large models or ensembles, e.g., pruning, transfer learning, or quantization. Model compression allows to transfer the function learned by a teacher ensemble or a single large discriminative model into a compact, faster student model exhibiting comparable performance \cite{DBLP:conf/kdd/BucilaCN06}. Instead of training the student model directly on a hand-labeled categorical dataset, this method employs a pre-trained teacher model to re-label the dataset and then train the compact neural network on this teacher-labeled dataset, using the raw predictions as the target. This training framework was shown to yield efficient models which perform better than if they had been trained on the hand-labeled dataset in a variety of discriminative tasks \cite{DBLP:conf/kdd/BucilaCN06, DBLP:conf/nips/BaC14, DBLP:conf/interspeech/LiZHG14}. Model compression was further extended and formalized into the general Knowledge Distillation (KD) framework \cite{DBLP:journals/corr/HintonVD15}. 


KD has been extensively applied in various fields and with other ends than model compression \cite{DBLP:conf/iclr/PapernotAEGT17, DBLP:conf/iclr/AnilPPODH18, DBLP:journals/tmm/YuanP20}. An interesting line of research that is closely related to ours proposes cross-task KD from image captioning and classification systems into an image synthesis generative neural-network \cite{DBLP:conf/mm/YuanP18, DBLP:journals/tmm/YuanP20}. In audio, KD was extensively used on Automatic Speech Recognition (ASR) tasks in order to exploit large unlabelled datasets \cite{DBLP:conf/interspeech/LiZHG14}, distill the knowledge from deep Recurrent Neural Networks (RNN) \cite{DBLP:conf/interspeech/ChanKL15} or, inversely, to improve the performance of deep RNN models by distilling knowledge from simple models as a regularization technique \cite{DBLP:conf/icassp/TangWZ16}. Works related to ours use KD as a means to adapt a model to a different audio domain task \cite{DBLP:conf/icassp/AsamiMYMA17} or even data modality (by distilling knowledge from a video classifier) \cite{DBLP:conf/nips/AytarVT16}, where labeled datasets are scarce, and large models would easily overfit. Some works employ KD to fuse knowledge from different audio representations into a single compact model \cite{DBLP:journals/corr/abs-2002-09607}. 


\vspace{-.2cm}
\section{Background}\label{sec:background}
\vspace{-.2cm}
This section provides a brief introduction to dark knowledge and explains the general knowledge distillation framework. 

\subsection{Dark Knowledge}
In the seminal work on Knowledge Distillation (KD) \cite{DBLP:journals/corr/HintonVD15}, the authors demonstrate that the improved performance of smaller models is due to the implicit information existent in the teacher's output probabilities (i.e., soft labels). As opposed to hard labels,  soft labels contain probability values for all of the output classes. The relative probability values that a specific data instance takes for each class contain information about how the teacher generalized the discriminative task. This hidden information existent in the relative probability values was termed \emph{dark knowledge} \cite{dark_knowledge}. An interesting observation on dark knowledge is that in KD, the student model can learn to correctly classify categories even if the training set does not contain examples thereof \cite{DBLP:journals/corr/HintonVD15}. In DarkGAN, we test if this principle can be transferred to audio generation. Many of the AudioSet attributes are not directly linked with the actual training data (e.g., the attributes "reverberation", "meow", or "drum" have little or no relationship to the tonal instrument sounds of the NSynth dataset). However, we hope that the implicit dark knowledge existent in the teacher-labeled data can help DarkGAN learn a coherent feature control over such attributes.

\subsection{Knowledge Distillation}\label{sec:dkd}
Multi-label classifiers typically produce a probability distribution over a set of classes by using a \emph{sigmoid} output layer that converts the so-called logit (the NN output before the activation function), \(z_i\), computed for the \(ith\) class into a probability \(q_i\) as 

\begin{equation}\label{eq:sigmoid}
    q_i =\frac{1}{1 + e^{-\frac{z_i}{T}}},
\end{equation}
where \(T\) is a temperature that is typically set to 1. In KD, knowledge is transferred to the distilled model by training it on the teacher-labeled data, using a higher temperature. By that, the distribution gets "compressed," emphasizing lower probability values. The same (higher) temperature is used while training the distilled model, but the temperature is set back to 1 after training. As for cost function, the binary cross-entropy is used as
\begin{equation}
   H_s(q) = - \frac{1}{N}\sum_{i=1}^{N} p_i \log{(q_i)} + (1-p_i)\log{(1 - q_i)},
\end{equation}
where \(N=128\) is the number of attributes, \(p_i\) are the soft-labels predicted by the teacher, and \(q_i\) is the probability predicted by the student model for the \(ith\) class.

\section{Experiments}\label{sec:experiments}

In this section, details are given about the conducted experiments. We describe the AudioSet ontology, the teacher and student architectures,  the metrics employed for evaluation, and the baselines used for comparison.
\subsection{Dataset}

We employ a subset from NSynth \cite{wavenetae} for our experiments. NSynth contains approximately 300k  single-note audios played by more than 1k different instruments from 10 different families. Each sample's onset occurs at time 0. The dataset contains various labels (e.g., pitch, velocity, instrument type), but we only use (i.e., condition the model on) pitch information in this work.  Each sample is four seconds long, with a 16kHz sample rate. For computational simplicity, we use only the first second of each sample. Also, we only consider samples with a MIDI pitch range from 44 to 70 (103.83 - 466.16 Hz), resulting in a subset of approximately 90k sounds equally distributed across the pitch classes. For the evaluation, we perform a 90/10\% split of the data.

Previous works on adversarial audio synthesis \cite{nistal1, gansynth} demonstrated that the Magnitude and Instantaneous Frequency of the STFT works well as a representation for harmonic sounds. We use an FFT size of 2048 bins and an overlap of 75\%.

\subsection{The AudioSet Ontology}\label{sec:aset}

AudioSet \cite{DBLP:conf/icassp/GemmekeEFJLMPR17} is a large-scale dataset containing audio data and an ontology of sound events that seek to describe real-world sounds. It was created to set a benchmark in the development of automatic audio event recognition systems, similar to those in computer-vision, such as ImageNet \cite{imagenet}. The dataset consists of a structured vocabulary of 632 audio event classes and a collection of approximately 2M human-labeled 10-second sound clips drawn from YouTube videos. The ontology is specified as a hierarchy of categories with a maximum depth of 6 levels, covering a wide range of human and animal sounds, musical genres and instruments, and environmental sounds. We encourage the reader to visit the corresponding website for a complete description of the ontology.\footnote{\url{research.google.com/audioset/ontology/}}

In this work, we do not employ all of the AudioSet attributes, as many of them refer to properties that are too vague for musical sounds or describe broader time-scale aspects of the sound (e.g., music, chatter, sound effect). Instead, we rank the attributes based on the geometric mean of their $90^{th}$ percentile (calculated on the predicted class probabilities for each attribute across the dataset), and the teacher's reported accuracy as $\sqrt{p^i_{90_{th}}\times acc^{i}}$. Then, we take the first 128 attributes according to this ranking.



\subsection{Models}

In the following, we introduce the teacher model and DarkGAN's architecture.

\subsubsection{Pre-trained AudioSet Classifier}
In this work, we distill the knowledge from a
pre-trained audio-tagging neural network (PANN) trained on raw audio recordings from the AudioSet collection \cite{DBLP:journals/taslp/KongCIWWP20}. PANNs were originally proposed for transferring knowledge to other audio
pattern recognition tasks. However, we use them to transfer the knowledge to a generative model and steer the generation process through a comprehensible vocabulary of attributes.

We employ the \emph{CNN-14} model from the PANNs \cite{DBLP:journals/taslp/KongCIWWP20}. \emph{CNN-14} is built upon a stack of 6 convolution-based blocks containing 2 CNN layers with a kernel size of 3x3. Batch Normalization is applied after every convolutional layer, and a ReLU non-linearity is used as the activation function. After each convolutional block, they apply an average-pooling layer of size 2x2 for down-sampling. Global pooling is applied after the last convolutional layer
to summarize the feature maps into a fixed-length vector. An extra fully-connected layer is added to extract embedding features before the output Sigmoid activation function. For more details on the architecture, please refer to \cite{DBLP:journals/taslp/KongCIWWP20}.

\subsubsection{DarkGAN}

\begin{figure}[t]
    \centering
    \includegraphics[scale=0.4, width=0.93\columnwidth]{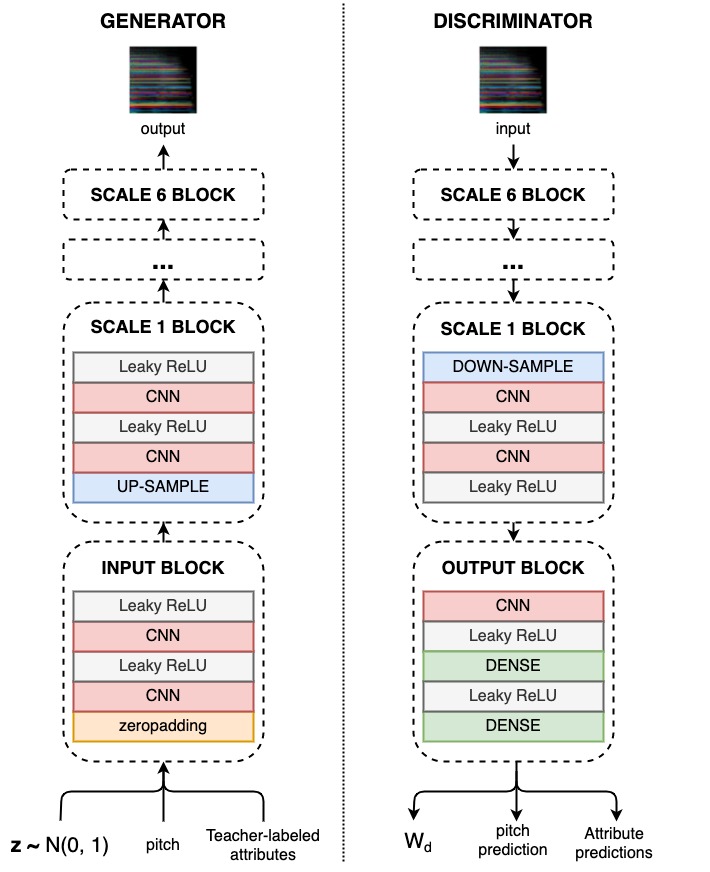}
     \vspace{-0.4cm}
    \caption{Proposed architecture for DarkGAN \cite{nistal2}}
    \label{fig:arch}
    \vspace{-0.4cm}
\end{figure}

The proposed GAN architecture, illustrated in Fig.~\ref{fig:arch}, follows the architecture of DrumGAN \cite{nistal2}. The input to $G$ is a concatenation of 128 teacher-labeled AudioSet attributes \(\alpha \in [0, 1]^{128}\) (see Sec.~\ref{sec:aset}), a one-hot vector $p \in \{0, 1\}^{26}$ containing the pitch class, and a random vector $z \sim \mathcal{N}_{32}(0,1)$. The resulting vector is placed as a column in the middle of a 4D tensor with $128+32+26$ convolutional maps. Then, it is fed through a stack of convolutional and box up-sampling blocks to generate the output signal \(x=G_\theta(z, p, \alpha)\). The number of feature maps decreases from low to high resolution as \{256, 128, 128, 128, 128, 64\}. The discriminator \(D\) mirrors $ G$'s configuration and estimates the Wasserstein distance $W_d$ between the real and generated distributions \cite{Gulrajani2017}, and predicts the AudioSet features accompanying the input audio in the case of a real batch, or those used for conditioning in the case of generated audio. In order to promote the usage of the conditioning information by $G$, we add to the objective function an auxiliary binary cross-entropy loss term for the distillation task and a categorical cross-entropy for the pitch classification task \cite{OdenaOS17}. 



\vspace{-0.1cm}
\subsection{Evaluation}\label{sec:eval}
\vspace{-0.1cm}
The task of synthesizing perceptually realistic audio is hard to formalize. In conditional models, as is the case in this work, an additional challenge is to assess whether the model is soundly responsive to the conditional input. In order to evaluate these properties of our model, a diverse set of objective metrics are computed. We compute these metrics for DarkGAN when trained under different temperature values in the distillation process (see Sec.~\ref{sec:kd}), as well as for various baselines. In this section, we describe these metrics as well as the baselines used for comparison.

\subsubsection{Scores and distances}
Following previous methodology \cite{nistal1, nistal2, DBLP:journals/corr/abs-2103-07390}, we compare real and generated distributions employing these metrics:
\begin{itemize}
    
    \item \textbf{Inception Score (IS)} \cite{SalimansGZCRCC16} penalizes models whose samples cannot be reliably classified into a single class or that only belong to a few from all possible classes. 
    We report on the Pitch Inception Score (PIS) and the Instrument Inception Score (IIS) \cite{nistal1}.
    
    \item \textbf{Kernel Inception Distance (KID)} \cite{BinkowskiSAG18} measures the dissimilarity between embeddings of real and generated samples. A low KID means that the generated and real distributions are close to each other.
    
    \item \textbf{Fréchet Audio Distance (FAD)} \cite{fad} measures the distance between continuous multivariate Gaussians fitted to embeddings of real and generated data. The lower the FAD, the smaller the distance between distributions of real and generated data.
\end{itemize}

\subsubsection{Consistency of Attribute Controls}\label{sec:cons}
This work aims to learn semantically meaningful controls with DarkGAN by distilling knowledge from an audio-tagging system trained on attributes from the AudioSet ontology. Therefore, we evaluate if changing an input attribute is reflected in the corresponding output of DarkGAN. To that end, we examine the change in the prediction of the teacher model (w.r.t. the output of DarkGAN) when changing a particular DarkGAN input attribute. A second property to assess is whether the \emph{dark knowledge} helps DarkGAN learn well-formed representations of specific attributes and generalize to out-of-distribution input combinations. To assess these two aspects, we perform the following tests:
\begin{enumerate}
    \item \emph{Attribute correlation}: we generate 10k samples using attribute vectors from the validation set as input to DarkGAN. The generated samples are fed to the teacher model to predict the attributes again. Then, for each attribute $i$, we compute the correlation between the input vector $\alpha$ and the predictions $\hat{\alpha}$ as
    \begin{equation*}
        \footnotesize
         \rho^i(\hat{\alpha}, \alpha) = \rho(F^i(G(z, p, \alpha)), \alpha),
    \end{equation*}
   where $F^i$ is the classifier's prediction for the $i$th attribute, $p$ is the pitch, and $z$ is the random noise.
    
    \item \emph{Out-of-distribution Attribute Correlation}: for each attribute $i$ exhibiting a positive correlation, i.e., $\mathcal{S} = \{\rho^i:\rho^i>0$\}, test (1) is repeated 50 times, but using 1k samples instead of 10k. In each repetition, a specific attribute is progressively incremented by an amount $\delta_l := 10^{-3 + l\frac{3.6}{50}}, l = 0, 1, ..., 50$\renewcommand{\thefootnote}{*}\footnote{The step of $\delta_l$ is defined to obtain more density of points in the range of variation of the attributes (i.e., [0, 1]) as well as $\delta_l>1$.} and we calculate
    \begin{equation*}
    \footnotesize
        \overline{\rho}_{\delta_l} = \frac{1}{\mid \mathcal{S} \mid}\sum_{\mathcal{S}} \rho^i(\hat{\alpha}, \alpha + \delta_l).
    \end{equation*}

    \item \emph{Increment consistency}: for the 50 attributes with the highest correlation, we compute
    \begin{equation*}
    \footnotesize
       \overline{\Delta F_{\delta_k}}= \sum_{i=1}^{50}\sum_{j=1}^{100}\frac{F^i(G(z_j, p_j, \alpha_j + \delta_k)) - F^i(G(z_j, p_j, \alpha_j))}{50\times100\times std(F^i (G(\mathbf{z}, \mathbf{p}, \boldsymbol{\alpha})))},
    \end{equation*}
    where $\alpha_j$ is the $j$th original feature vector from a set of 100 samples randomly picked from the validation set, and $\delta_k := \frac{k}{5}, k = 0, 1, ..., 25$. Intuitively, it is defined as the average difference of the predicted attributes of the generated audios (i.e., the difference before and after the attribute increment) as a function of the increment $\delta_k$. We express the result in terms of standard deviations of the non-incremented generated examples as $\text{std}(G(z, p, \alpha))$.
\end{enumerate}

\subsubsection{Baselines}\label{sec:baselines}
We compare the metrics described above with \emph{real data} to obtain a baseline for each metric. Also, GANSynth \cite{gansynth}, the state-of-the-art on audio synthesis with GANs, is used for comparison.\footnote{\scriptsize \url{https://github.com/magenta/magenta/tree/master/magenta/models/gansynth}} As GANSynth generates 4-second long sounds, the waveform is trimmed down to 1 second for comparison with our models. Additionally, we examine the effect that KD has on these metrics by comparing against a model analogous to DarkGAN, but without using the AudioSet feature conditioning (\emph{baseline}). Experiment results for DarkGAN are shown for different temperature values $T \in \{1, 1.5, 2, 3, 5\}$ (\ref{eq:sigmoid}) as part of the KD process (see Sec.~\ref{sec:dkd}), and we report separate results for conditional attributes obtained from the training (tr) and validation (val) set.

\vspace{-0.3cm}
\section{Results}\label{sec:discussion}
\vspace{-0.1cm}

In this section, we present the results from the evaluation procedure described in Sec.~\ref{sec:eval}. Furthermore, we validate the quantitative results based on an informal assessment of the generated content. 
\vspace{-0.1cm}
\subsection{Quantitative Metrics}\label{sec:res1}
Table~\ref{tab:metrics} presents the metrics scored by DarkGAN$_T$ and the baseline models, as described in Sec.~\ref{sec:baselines}. Note that we condition DarkGAN on attribute vectors randomly sampled from the validation set. Overall, DarkGAN$_{T \in \{1.5, 2\}}$ obtains better results than the baselines and is close to \emph{real data} in most metrics. All models score higher PIS than \emph{real data}, with GANSynth in the first place, suggesting that the generated examples have a clear pitch and that the distribution of pitch classes follows that of the training data. This is not surprising, as all the models have explicit pitch conditioning. In contrast, we do not provide conditioning attributes for the instrument class. Therefore, we observe a slight drop in IIS for all models compared to \emph{real data}. DarkGAN$_{T \in \{1.5, 2\}}$ achieves the highest IIS, suggesting that the model captured the timbre diversity existent in the dataset and, also,  that the generated sounds can be reliably classified into one of all possible instruments. In terms of KID, DarkGAN$_{T \in \{1.5, 2\}}$ and \emph{baseline} are on a par with \emph{real data}. A KID equal to \emph{real data} indicates that the Inception embeddings are similarly distributed for real and generated data. As our Inception classifier is trained on pitch and instrument classification and predicting AudioSet features, similarities in such an embedding space indicate common timbral and tonal characteristics between the generated and the real audio data distribution. This trend is maintained in the case of the FAD, where DarkGAN$_{T=2}$ obtains the best scores followed closely by DarkGAN$_{T \in \{1, 1.5\}}$.

From the results discussed above, we can conclude that distilling knowledge from the AudioSet classifier helps DarkGAN learning the real data distribution. Furthermore, using slightly higher temperatures in the distillation process yields an improvement over the \emph{baseline} without feature conditioning. We speculate that the additional supervised information that the teacher model provides to DarkGAN's discriminator results in a more meaningful gradient for the generator. Also, attribute conditioning (i.e., attribute vectors sampled from the validation set) may help the generator synthesize diverse samples closer to the training data distribution.

\begin{table}[]
    \centering
    \scriptsize
    \begin{tabular}{c|cc|cc|cc|cc}
         Model & \multicolumn{2}{c}{PIS$\uparrow$} & \multicolumn{2}{c}{IIS$\uparrow$} & \multicolumn{2}{c}{KID$\downarrow^{\mathrm{a}}$} & \multicolumn{2}{c}{FAD$\downarrow$}\\
         \toprule
         \emph{real data} & \multicolumn{2}{c}{17.7} & \multicolumn{2}{c}{5.7} & \multicolumn{2}{c}{6.7} & \multicolumn{2}{c}{0.1} \\
         GANSynth \cite{gansynth} & \multicolumn{2}{c}{\textbf{19.6}} & \multicolumn{2}{c}{4.0} & \multicolumn{2}{c}{7.1} & \multicolumn{2}{c}{4.5} \\
         \emph{baseline} & \multicolumn{2}{c}{18.5} & \multicolumn{2}{c}{4.3} & \multicolumn{2}{c}{\textbf{6.7}} & \multicolumn{2}{c}{0.8} \\
         \midrule
          DarkGAN$_{T}$ & tr & val & tr & val & tr & val & tr & val\\
         \midrule
         $T=1$ & 18.4 & 18.3 & 4.0 & 4.0 & 6.8 & 6.8 &  0.7 & 0.7\\
         $T=1.5$ & 19.0 & 19.0 & \textbf{4.5} & \textbf{4.5} & \textbf{6.7} & \textbf{6.7} & 0.7 & 0.7\\
         $T=2$ & 19.1 & 19.0 & 4.2 & 4.1 & \textbf{6.7} & 6.8 & \textbf{0.6} & \textbf{0.6}\\
         $T=3$ & 19.1 & 19.1 & 4.2 & 4.1 & 6.8 & 6.8 & 0.8 & 0.8\\
         $T=5$ & 19.2 & 19.1 & 4.0 & 4.0 & 6.8 & 6.8 & 0.8 & 0.8\\
         \bottomrule
         \multicolumn{9}{l}{$^{\mathrm{a}}$\(\times{10^{-4}}\)}
         \vspace{-.2cm}
    \end{tabular}
    \caption{PIS, IIS, KID and FAD (see Sec.~\ref{sec:eval})}
    \label{tab:metrics}
\end{table}


\vspace{-0.1cm}
\subsection{Attribute Consistency and Generalisation}\label{sec:res2}
\vspace{-0.1cm}

\begin{table}[]
    \scriptsize
    \centering
    \begin{tabular}{c|ccccc}
         Attribute & T=1 & T=1.5 & T=2 & T=3 & T=5\\
         \toprule
         Acoustic guitar & 0.20 & 0.36 & \textbf{0.39} & 0.23 & 0.10\\
         Bass guitar & 0.30 & 0.38 & \textbf{0.46} & 0.38 & 0.19\\
         Brass Instrument & 0.28 & \textbf{0.49} & 0.38 & 0.26 & 0.00\\
         Cello & 0.24 & \textbf{0.29} & 0.26 & 0.17 & 0.00\\
         Chime & 0.15 & 0.33 & \textbf{0.39} & 0.31 & 0.03\\
         Guitar & 0.28 & 0.37 & \textbf{0.42} & 0.34 & 0.13\\
         Plucked string & 0.27 &  0.37 & \textbf{0.42} & 0.32 & 0.11\\
         Saxophone & 0.25 & \textbf{0.41} & \textbf{0.41} & \textbf{0.41} & 0.03\\
         Trombone & 0.18 & \textbf{0.41} & 0.29 & 0.16 & 0.00\\
         Trumpet & 0.16 & \textbf{0.46} & 0.36 & 0.25 & 0.00\\
         ... & \multicolumn{5}{c}{...}\\
        Didgeridoo & 0.06 & 0.16 & \textbf{0.21} & 0.20 & 0.08\\
        Drum & 0.05 & 0.21 & \textbf{0.24} & 0.12 & 0.01\\
        Electronic tuner & 0.35 & 0.44 & \textbf{0.50} & 0.29 & 0.13\\
        Percussion & 0.04 & 0.19 & \textbf{0.30} & 0.14 & 0.08\\
        Sine wave & 0.28 & \textbf{0.32} & 0.27 & 0.17 & 0.10\\
        Singing bowl & 0.08 & 0.20 & \textbf{0.24} & 0.21 & 0.03\\
        Siren & 0.13 & 0.19 & \textbf{0.24} & 0.10 & 0.08\\
        Tuning fork & 0.22 & 0.29 & \textbf{0.35} & 0.29 & 0.10\\
        Zither & 0.03 & 0.18 & \textbf{0.19} & 0.07 & -0.01\\
         ... & \multicolumn{5}{c}{...}\\
         Cat & -0.01 & -0.01 & -0.01 & -0.01 & 0.00\\
         Chicken, rooster & 0.00 & -0.06 & -0.02 & -0.01 & -0.01\\
         Domestic animals, pets & -0.01 & -0.02 & -0.02 & 0.00 & 0.00\\
         Frog & 0.00 & 0.03 & 0.07 & 0.06 & -0.03\\
         Insect & 0.00 & -0.02 & -0.02 & -0.02 & -0.01\\
         Speech & -0.04 & -0.10 & -0.07 & -0.05 & 0.01\\
        \bottomrule
    \end{tabular}
    \caption{A few examples of attribute correlation coefficients $\rho^i(\hat{\alpha}, \alpha)$ (see Sec.~\ref{sec:cons}).}
    \label{tab:corr}
    \vspace{-0.5cm}
\end{table}

Note that the metrics discussed in this section are not guaranteed to relate directly to human perception, but we consider them suitable indicators of whether the model responds coherently to the input conditioning. There exists the threat of the generator producing adversarial examples, but we argue that this is prevented by the discriminator having to satisfy the Wasserstein criterion (as adversarial examples would exhibit out-of-distribution artifacts). This assumption is also supported by informal listening tests where we find that the metrics correlate with our perception (see Sec.~\ref{sec:informal}).

Table~\ref{tab:corr} shows the results for the \emph{attribute correlation} $\rho^i(\hat{\alpha}, \alpha)$ (see Sec.~\ref{sec:cons}). At the top of the table, we show a few attributes corresponding to classes represented in the NSynth dataset (e.g., "guitar", "trumpet"). In the middle, we show attributes that, while not being present in the dataset (e.g., "siren", "tuning fork"), still exhibit (relatively) high correlation. At the bottom, attributes that obtain low correlations are presented (e.g., "cat", "insect"). We can observe that models trained with $T \in \{1.5, 2, 3\}$ generally obtain better results than $T \in \{1, 5\}$ in most attributes. Specifically, DarkGAN$_{T=2}$ yields the highest correlations, followed by DarkGAN$_{T=1.5}$. Note that temperatures higher than 1 also improve the correlation for attributes that do not have corresponding classes in the dataset (e.g., "didgeridoo", "percussion", "singing bowl"). This suggests that DarkGAN can extract dark knowledge (which is emphasized by increasing $T$) from the soft labels. The soft labels indicating the presence of (potentially just slight) timbral characteristics in various sounds are helping the model to learn linearly dependent feature controls for those attributes.

A more in-depth analysis of feature errors and the distribution of features in the dataset would be required to further characterize the results for each attribute. However, it is reasonable that those classes obtaining higher correlations share some timbral features with the training data (e.g., clearly, "violins" are contained in the data set, and a "tuning fork" is similar to a "mallet"). In contrast, those attributes obtaining low correlations may be related to underrepresented features in the training set or features that the model failed to capture.

\begin{figure}[htb]
\vspace{-0.5cm}
\centering
\begin{tikzpicture}
  \node (img1)  {\includegraphics[scale=0.42, trim={1cm 0.7cm 1.6cm 1cm},clip]{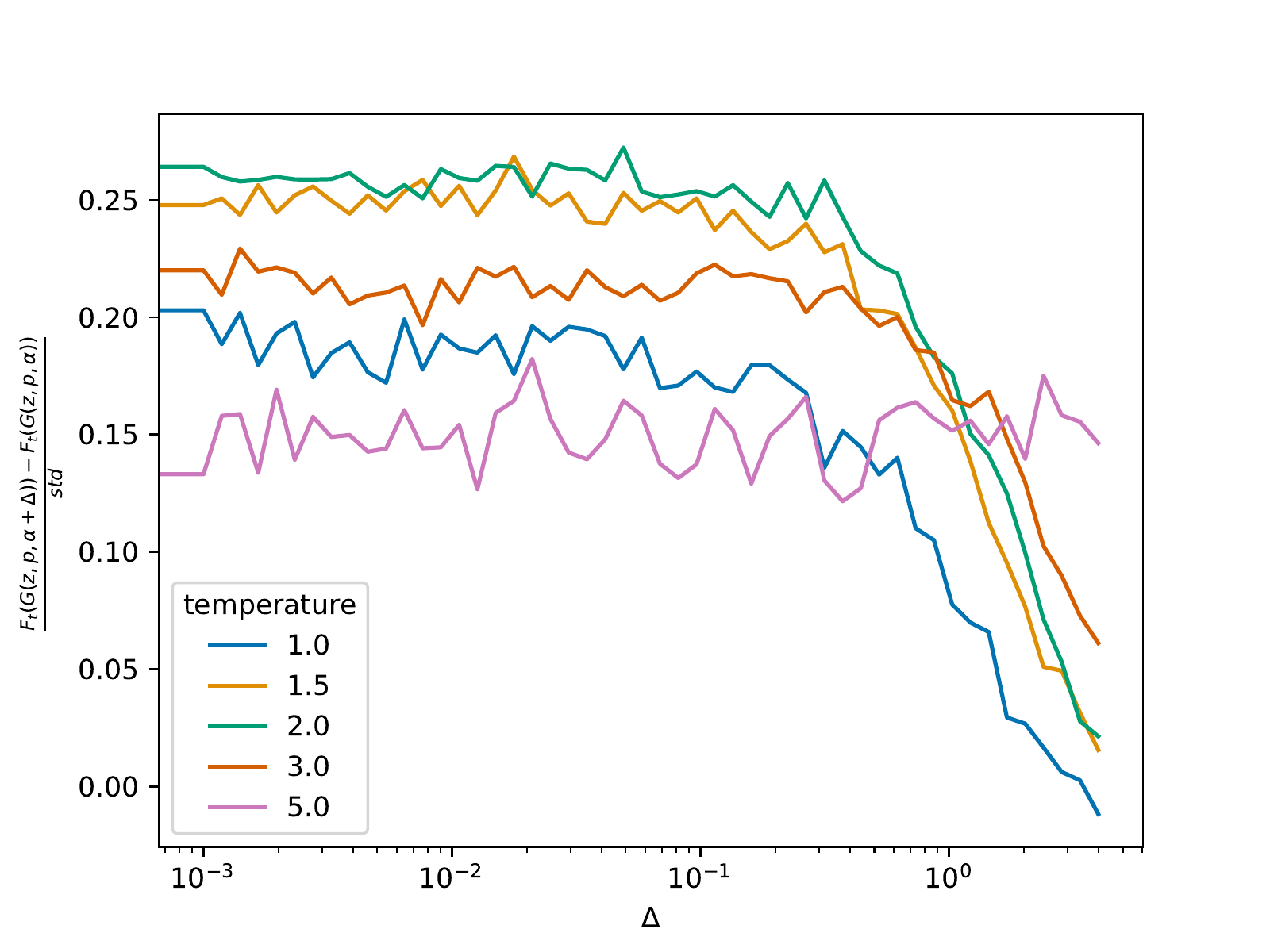}};
  \node[below=of img1, node distance=0cm, yshift=1cm]{$\delta_l$};
  \node[left=of img1, node distance=0cm, rotate=90, anchor=center,yshift=-0.7cm] {$\overline{\rho}_{\delta_l}$};
\end{tikzpicture}
\vspace{-0.5cm}
\caption{Out-of-distribution average attribute correlation $\overline{\rho}_{\delta_l}$ (see Sec.~\ref{sec:cons})}
\label{fig:corr}
\end{figure}

Fig.~\ref{fig:corr} shows the correlation coefficient when increasing each attribute by a value $\delta_l$ in the input conditioning. The plot reveals that the trend of Table~\ref{tab:corr} is maintained throughout an ample range of variation of the attributes. Interestingly, while the correlation of DarkGAN$_{T=1}$ considerably declines after an increase $\delta_l > 10^{-0.8}$, using a temperature ${T \in \{1.5, 2, 3\}}$ the decline is more moderate, and we observe some correlation even for a $\delta_l > 1$, which is outside the range of the attributes. 

\begin{figure}[htb]
\vspace{-0.5cm}
\centering
\begin{tikzpicture}
  \node (img1)  {\includegraphics[scale=0.42, trim={1.1cm 0.7cm 1.6cm 1cm},clip]{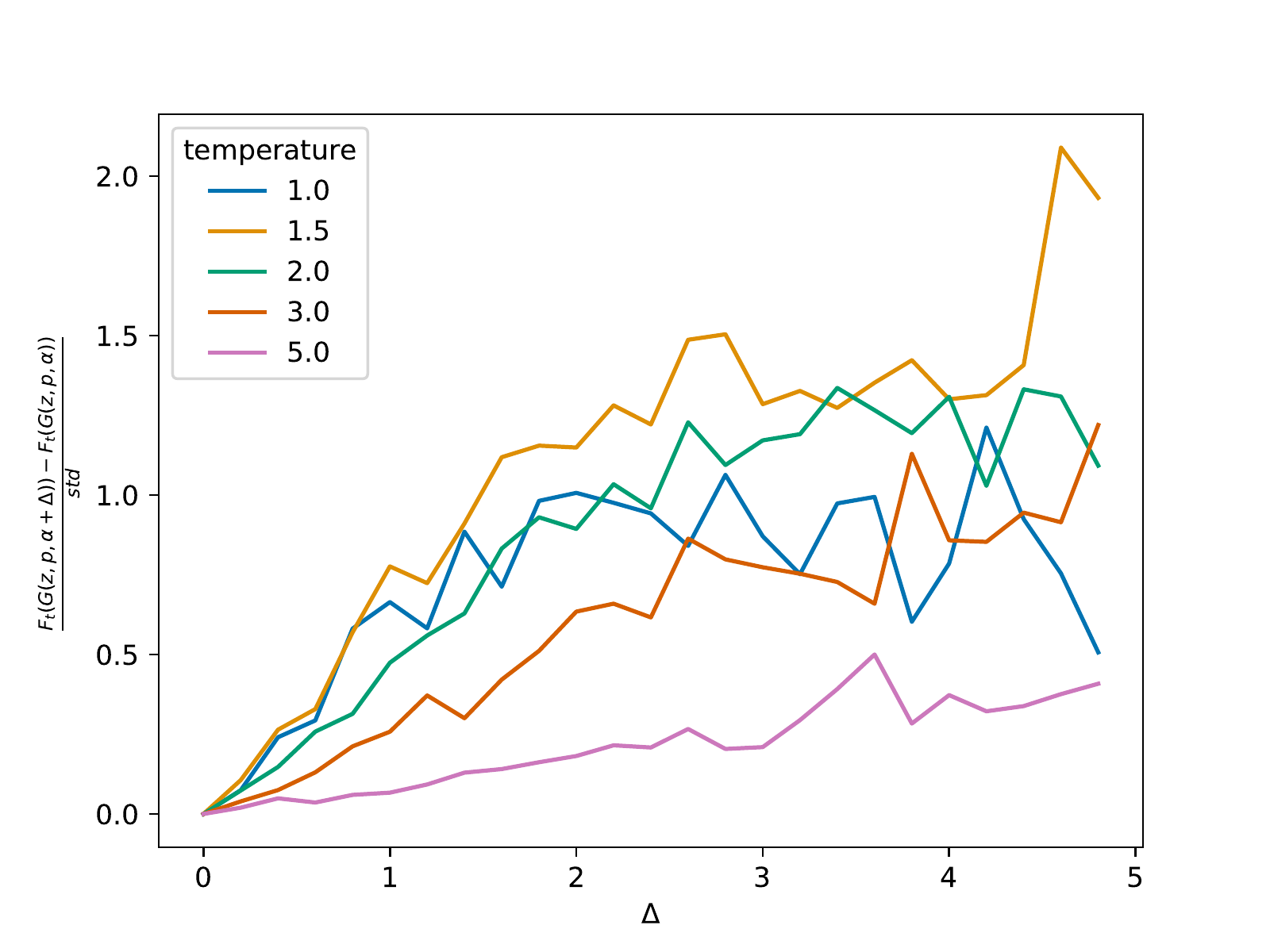}};
  \node[below=of img1, node distance=0cm, yshift=1cm]{$\delta_k$};
  \node[left=of img1, node distance=0cm, rotate=90, anchor=center,yshift=-0.7cm] {$\overline{\Delta F}_{\delta_k}$};
\end{tikzpicture}
\vspace{-0.5cm}
\caption{Increment consistency $\overline{\Delta F}_{\delta_k}$ (see Sec.\ref{sec:cons})}
\vspace{-.5cm}
\label{fig:increment}
\end{figure}

As the correlation coefficient provides normalized results (regarding scale and offsets), we evaluate the attribute control using the \emph{increment consistency} metric $\overline{\Delta F}_{\delta_k}$ (see Fig.~\ref{fig:increment}). 
We observe that for low increments of the features ($\delta_k < 1$) temperatures $T \in \{1, 1.5, 2\}$ yield comparable input-output relationships of the features. A temperature $T=1.5$, however, yields more consistent feature differences for increments $\delta_k > 1$ of the conditional input features. 
In conclusion, while $DarkGAN_{T=2}$ yields better correlation over all the data (i.e., conditional and predicted attributes are more strongly dependent), for attributes with particularly high correlation, $DarkGAN_{T=1.5}$ performs best in over-emphasizing dark knowledge contained in the data (i.e., the degree of change is higher, especially for $\delta_k > 1$).

\vspace{-.1cm}
\subsection{Informal Listening}\label{sec:informal}
In the accompanying website,\footnote{https://an-1673.github.io/DarkGAN.io/} we show sounds generated under various conditioning settings, including generations with feature combinations randomly sampled from the validation set, generations where we fix $\alpha$ and $p$ while changing $z$, timbre transfer, scales, and more. Overall, we find the results of PIS, IIS, KID, and FAD, discussed in Sec.~\ref{sec:res1}, to align well with our perception. The quality of the generated audio is acceptable for all models. Also, we find the generated examples to be diverse in terms of timbre, and the tonal content is coherent with the pitch conditioning. Moreover, we perceive that most of the attributes exhibiting high correlations (see Table~\ref{tab:corr}) are audible in the generated output, particularly in the case of DarkGAN$_{T \in \{1, 1.5, 2\}}$. For higher temperatures $T \in \{3, 5\}$, the model's responsiveness to the attribute conditioning drops substantially. We find the model to be particularly responsive to attributes such as "drum", "tuning fork", "theremin", "choir", or "cowbell". To other attributes (e.g., "accordion", "piano", or "organ"), even though the analysis yields moderate correlations, the model does not seem to produce perceptually satisfactory outputs.

\vspace{-0.3cm}
\section{Conclusion}\label{sec:conclusion}
\vspace{-0.1cm}
In this work, we distilled knowledge from a large-scale audio tagging system into DarkGAN, an adversarial synthesizer of tonal sounds. The goal was to enable steering the synthesis process using attributes from the AudioSet ontology. A subset of the NSynth dataset was fed to a pre-trained audio tagging system to obtain AudioSet predictions. These predictions were then used to condition DarkGAN. The proposed Knowledge Distillation (KD) framework was evaluated by comparing different temperature settings and employing a diverse set of metrics. Results showed that DarkGAN can generate audio resembling the true dataset and enables moderate control over a comprehensible vocabulary of attributes. By slightly increasing the temperature during the distillation process, we can further improve the responsiveness of the attribute controls. It is also notable that KD can be performed even when the original dataset (i.e., the AudioSet collection) is not involved.

\section{Acknowledgements}
This research is supported by the European Union's Horizon 2020 research and innovation program under the Marie Sklodowska-Curie grant agreement No.~765068 (MIP-Frontiers).

\bibliography{main}

%
%
%
%
%

\end{document}